\documentclass[prd,aps,twocolumn,a4paper,superscriptaddress,floatfix,showpacs,nofootinbib]{revtex4-1}

\usepackage{graphicx,psfrag}
\usepackage{amsmath,amsfonts,amssymb}
\usepackage{mathrsfs}
\usepackage{multirow,hyperref}
\usepackage{comment}
\usepackage{pifont}
\usepackage[normalem]{ulem}
\usepackage{scrextend}
\usepackage[dvipsnames]{xcolor}

\newcommand{\be}{\begin{equation}}
\newcommand{\ee}{\end{equation}}
\newcommand{\bsube}{\begin{subequations}}
\newcommand{\esube}{\end{subequations}}

\def\({\left(}
\def\){\right)}

\def\Msun{M_\odot}

\usepackage{color}

\begin{document}

\title{How loud are neutron star mergers?}

\author{Sebastiano \surname{Bernuzzi}}
\affiliation{DiFeST, University of Parma, and INFN, I-43124
  Parma, Italy}
\affiliation{TAPIR, Walter Burke Institute for Theoretical Physics, 
  California Institute of Technology, 1200 E 
California Blvd, Pasadena, California 91125, USA}
\author{David \surname{Radice}}
\affiliation{TAPIR, Walter Burke Institute for Theoretical Physics, 
  California Institute of Technology, 1200 E 
California Blvd, Pasadena, California 91125, USA}
\author{Christian D. \surname{Ott}}
\affiliation{Yukawa Institute for Theoretical Physics, Kyoto University, Kyoto, Japan}
\affiliation{TAPIR, Walter Burke Institute for Theoretical Physics, 
  California Institute of Technology, 1200 E 
  California Blvd, Pasadena, California 91125, USA}
\author{Luke F. \surname{Roberts}}
\affiliation{TAPIR, Walter Burke Institute for Theoretical Physics, 
  California Institute of Technology, 1200 E 
California Blvd, Pasadena, California 91125, USA}
\affiliation{NASA Einstein Fellow}
\author{Philipp \surname{M\"osta}}
\affiliation{Department of Astronomy, University of California at
  Berkeley, 501 Campbell Hall 3411, Berkeley, California 94720, USA}
\affiliation{TAPIR, Walter Burke Institute for Theoretical Physics, 
  California Institute of Technology, 1200 E 
California Blvd, Pasadena, California 91125, USA}
\affiliation{NASA Einstein Fellow}
\author{Filippo \surname{Galeazzi}}
\affiliation{Institut f\"ur Theoretische Physik,
  Max-von-Laue-Stra{\ss}e 1, 60438 Frankfurt, Germany} 

\date{\today}

\begin{abstract}
We present results from the first large parameter
study of neutron star mergers using fully general relativistic
simulations with finite-temperature microphysical equations of
state and neutrino cooling. We consider equal and unequal-mass
binaries drawn from the galactic population and simulate each
binary with three different equations of state. Our focus is on the
emission of energy and angular momentum in gravitational waves in the
postmerger phase.  We find that the emitted gravitational-wave energy
in the first $\sim$$10\,\mathrm{ms}$ of the life of the resulting
hypermassive neutron star (HMNS) is about twice the energy emitted
over the entire inspiral history of the binary.  The total radiated
energy per binary mass is comparable to or larger than that of
nonspinning black hole inspiral-mergers. About 
  $0.8-2.5\%$ of the binary mass-energy is emitted at kHz frequencies
  in the early HMNS evolution. We find a clear dependence
    of the postmerger GW emission on binary configuration and equation
    of state and show that it can be encoded as a broad function of
    the binary tidal coupling constant $\kappa^T_2$.  Our results
  also demonstrate that the dimensionless spin of black holes resulting from
  subsequent HMNS collapse are limited to $\lesssim0.7-0.8$. This may
  significantly impact the neutrino pair annihilation mechanism for
  powering short gamma-ray bursts (sGRB).
\end{abstract}

\pacs{
  04.25.D-,    
  04.30.Db,    
  95.30.Sf,    
  95.30.Lz,    
  97.60.Jd     
}
\maketitle


\begin{figure*}[t]
  \begin{center}
  \includegraphics[width=\textwidth]{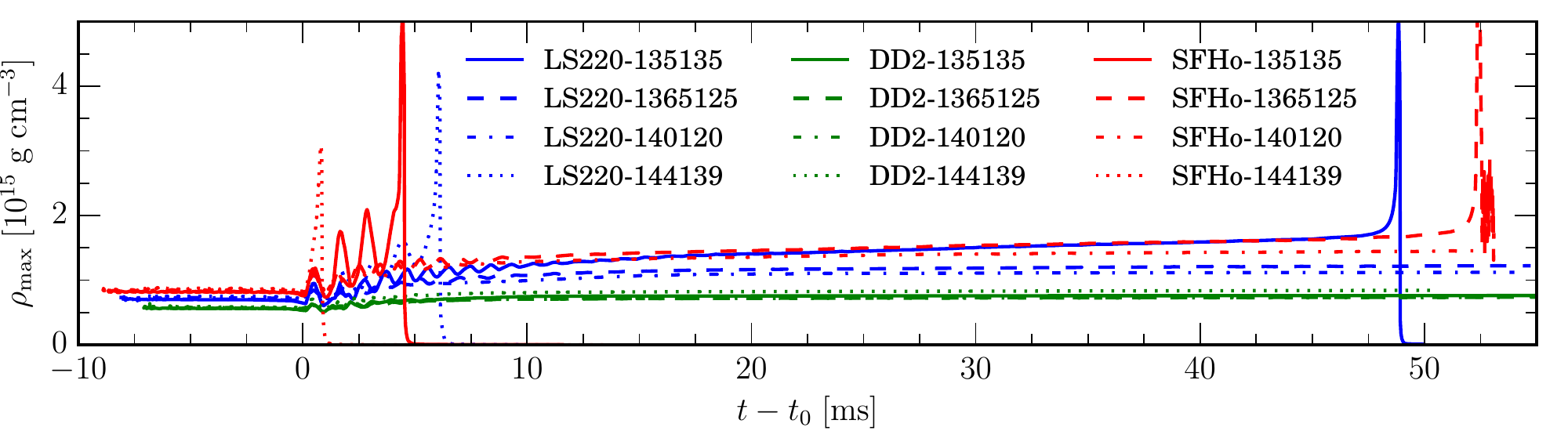}
  \end{center}
  \includegraphics[width=\textwidth]{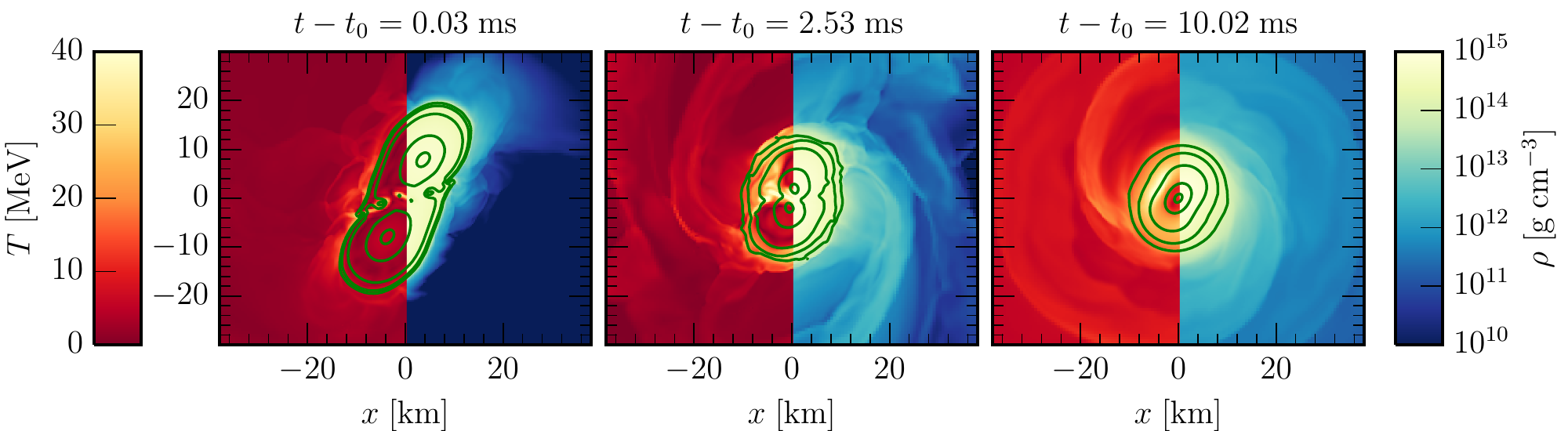}
  \caption{\emph{Top panel:} evolution of the maximum rest-mass density
  $\rho_{\max}$ for all the configurations. For simulations times $t<t_0$,
  $\rho_{\max}$ is the maximum value of the densest star; after contact and
  merger $\rho_{\max}$ is the absolute maximum. \emph{Bottom panel:} color coded
  temperatures and density at three representative times for LS220-135135. The
  black contours enclose densities larger than $10, 20, 40, 80$ and $98\ \%$ of
  $\rho_{\max}$. The core of the HMNS remains relatively cold, with $T \simeq 10\
  \mathrm{MeV}$ and is surrounded by a hot shell $T \simeq 40\ \mathrm{MeV}$ of
  material at densities $\sim$$5 \times 10^{14}\ \mathrm{g}\
  \mathrm{cm}^{-3}$.} 
  \label{fig:rhomax}
\end{figure*}

\section{Introduction}

Gravitational wave (GW) astronomy has been inaugurated by the first
direct detection of GWs from a binary black hole (BH) merger by 
Advanced LIGO \cite{Abbott:2016blz}. Another primary source for
Advanced LIGO is the GW-driven inspiral and 
merger of binary neutron stars (BNS).
A possible outcome of the merger is the formation of a hot,
differentially rotating hypermassive neutron star (HMNS), which may
survive for many tens of milliseconds before collapsing to
a BH,
e.g.~\cite{Rosswog:2001fh,*Rosswog:2003rv,Shibata:2006nm,Sekiguchi:2011zd,Bauswein:2011tp,Hotokezaka:2013iia,*Palenzuela:2015dqa}. Observations
of NSs with mass $\sim$$2\Msun$
\cite{Demorest:2010bx,Antoniadis:2013pzd} and of BNSs with individual
masses $\sim$$1.35\Msun$ \cite{Kiziltan:2013oja} favor the HMNS scenario
as the initial outcome. The stiff nuclear equation of
state (EOS) in combination with differential rotation at least
temporarily prevents collapse to a BH \cite{kaplan:14}.
GW emission is expected to depend on the interplay of several physical
ingredients: relativistic (magneto)hydrodynamics (M)HD, nonlinear
gravity, finite-temperature effects in the nuclear EOS, neutrino
cooling, and angular momentum redistribution (via viscosity or
(M)HD). Fully general relativistic (GR) simulations that include realistic
microphysics (i.e.~nuclear and neutrino physics) are the only reliable
means to study postmerger evolution and its GW emission.

In this work, we present results from a new and
largest-to-date set of BNS configurations simulated in full numerical
relativity with temperature-dependent microphysical EOS and neutrino
physics. Our configurations are representative of galactic BNS
systems. We consider three different EOS broadly consistent with
observational and experimental constraints. We focus on the
postmerger evolution and its GW emission, and show for the first time
that the HMNS phase is the most GW-luminous phase in the entire
history of BNS systems. Soft EOS and HMNS masses close to (but below)
the prompt collapse threshold are the most luminous. BHs
resulting from HMNSs that survive for $\gtrsim10\,\mathrm{ms}$ are
robustly limited to dimensionless spins $\lesssim0.7$. Larger spins
are obtained if the merger remnant collapses promptly or
within $1-2$ dynamical times of merger. 

\begin{table*}[t]
  \centering
  \caption{BNS properties (EOS, individual isolation masses, total baryonic mass
  of the binary, ADM quantities, dimensionless tidal coupling
  constant, e.g.~\cite{Bernuzzi:2014kca}) and the dimensionless
  radiated GW energy per 
  binary mass $E_\text{GW}/M$ and the mass-rescaled angular momentum $J/M^2$ at
  $t_{0}$ (merger) and $t_{N}$ ($N$~ms after merger). For
  configurations collapsing to a BH we also report $E^c_\text{GW}/M$ and $J^c/M^2$ as computed
  $\sim$$1\ \mathrm{ms}$ after collapse and the BH irreducibile mass and
  dimensionless angular momentum as measured by the horizon finder.
  All numbers are from simulations with $\Delta x=295$~m. The
  total binary mass is $M=M_A+M_B$. 
  Configurations are named according to EOS and masses $M_A$,$M_B$.}
  %
  \hspace{-1em}
  \begin{tabular}{l|cccccc|c@{\hskip 8pt}c@{\hskip 8pt}c@{\hskip 8pt}c|c@{\hskip 8pt}c@{\hskip 8pt}c@{\hskip 8pt}c|cc|cc}
    \hline
    \multirow{2}{*}{EOS} &
    $M_A$ & $M_B$ & $M_b$ & $M_\text{ADM}$ & $J_\text{ADM}$ & $\kappa^T_2$ &
    \multicolumn{4}{|c|}{$E_\text{GW}(t)/M\times10^{2}$} &
    \multicolumn{4}{|c|}{$J(t)/M^2 \times 10^1$} &
    $E^c_\text{GW}/M$ & $J^c/M^2$ &
    $M_\text{BH}$ & $a_\text{BH}$ \\
    & $[\Msun]$ & $[\Msun]$ & $[\Msun]$ & $[\Msun]$ & $[G\Msun^2/c]$ & &
    $t_0$ & $t_{10}$ & $t_{20}$ & $t_{50}$ &
    $t_0$ & $t_{10}$ & $t_{20}$ & $t_{50}$ &
    $\times 10^2$ & $\times 10^1$ &
    $[\Msun]$ & $\times 10^1$ \\
    \hline
    DD2
    & 1.40\phantom{0} & 1.20 & 2.829 & 2.576 & 6.537 & 203
    & 1.27 & 2.13 & 2.17 & 2.18 & 8.87 & 7.95 & 7.90 & 7.89 &  - & - & - & - \\
    DD2
    & 1.365 & 1.25 & 2.843 & 2.589 & 6.639 & 194
    & 1.34 & 2.24 & 2.29 & 2.31 & 8.87 & 7.91 & 7.86 & 7.83 & - & - & - & - \\
    DD2
    & 1.35\phantom{0} & 1.35 & 2.946 & 2.673 & 7.015 & 162
    & 1.37 & 2.56 & 2.58 & 2.60 & 8.75 & 7.57 & 7.54 & 7.53 & - & - & - & - \\
    DD2
    & 1.44\phantom{0} & 1.39 & 3.100 & 2.799 & 7.589 & 124
    & 1.46 & 2.90 & 2.95 & 2.97 & 8.60 & 7.29 & 7.25 & 7.23 & - & - & - & - \\
    \hline
    LS220
    & 1.40\phantom{0} & 1.20 & 2.830 & 2.574 & 6.540 & 159
    & 1.34 & 2.09 & 2.31 & 2.35 & 8.79 & 8.03 & 7.81 & 7.78 & - & - & - & -\\
    LS220
    & 1.365 & 1.25 & 2.846 & 2.588 & 6.623 & 151
    & 1.38 & 2.89 & 3.05 & 3.12 & 8.76 & 7.35 & 7.20 & 7.15 & - & - & - & - \\
    LS220
    & 1.35\phantom{0} & 1.35 & 2.947 & 2.671 & 7.000 & 125
    & 1.46 & 3.32 & 3.63 & - & 8.65 & 7.0 & 6.81 & - & 3.80 & 6.68 & 2.40 & 5.44 \\
    LS220
    & 1.44\phantom{0} & 1.39 & 3.102 & 2.797 & 7.570 & 94
    & 1.52 & - & - & - & 8.51 & - & - & - & 3.68 & 6.92 & 2.70 & 7.04 \\
    \hline
    SFHo
    & 1.40\phantom{0} & 1.20 & 2.850 & 2.573 & 6.525 & 115
    & 1.53 & 3.21 & 3.37 & 3.48 & 8.47 & 7.06 & 6.92 & 6.84 & - & - & - & - \\
    SFHo
    & 1.365 & 1.25 & 2.868 & 2.589 & 6.615 & 110
    & 1.52 & 3.61 & 3.80 & 3.94 & 8.47 & 6.78 & 6.63 & 6.53 & - & - & N.A. & N.A. \\
    SFHo
    & 1.35\phantom{0} & 1.35 & 2.972 & 2.674 & 7.018 & 89
    & 1.59 & - & - & - & 8.38 & - & - & -& 3.77 & 6.86 & 2.56 & 6.83 \\
    SFHo
    & 1.44\phantom{0} & 1.39 & 3.133 & 2.801 & 7.581 & 67
    & 1.66 & - & - & - & 8.26 & - & - & - & 2.27 & 7.86 & 2.79 & 8.08 \\
    \hline
  \end{tabular}
 \label{tab:bns}
\end{table*}

\section{Binary Configurations and Simulations}
The properties of the considered binary configurations are summarized
in Tab.~\ref{tab:bns}. We choose equal and unequal-mass configurations
guided by observed galactic BNS systems \cite{Kiziltan:2013oja}.
Configurations *-135135, *-136125, *-140120, and *-144139 
reproduce the NS masses in the binaries identified by B2127+11C (and
B1534+12), J1906+0746, J1756-2251 (and J1829+2456), and B1913+13,
respectively.  We simulate these binaries using three different
nuclear EOS, referred to as LS220 \cite{Lattimer:91}, DD2
\cite{Typel:10,*2010NuPhA.837..210H}, and SFHo \cite{Steiner:13b}.
They span a reasonable range of radii and maximum gravitational masses
for non-spinning NSs: DD2 has $M^\text{TOV}_\text{max}\sim$$2.42\Msun$
and radius $R_{1.35\Msun}\sim$$13.2$~km; SFHo and LS220 have similar
$M^\text{TOV}_\text{max}\sim$$2.05\Msun$, but $R_{1.35\Msun}\sim$$11.9$~km
(SFHo) and $R_{1.35\Msun}\sim$$12.7$~km (LS220).  We refer to EOS with
larger $R_{1.35\Msun}$ as being ``stiffer'', since at fixed mass, a
stiffer EOS results in lower central densities and larger NS radii.
All three EOS provide maximum cold NS masses greater than $2 \,
M_\odot$, which puts them in agreement with the maximum observed NS
mass \cite{Demorest:2010bx,Antoniadis:2013pzd}.  
SFHo and LS220 fall within the NS mass radius relation predicted by
\cite{Steiner:13}, while DD2 has a somewhat larger radius. SFHo and
DD2 both agree with microscopic neutron matter calculations
\cite{Fischer:14}, but LS220 falls outside of the favored region.

We compute conformally-flat initial data for our simulations, assuming
quasicircular orbits and irrotational flow \cite{gourgoulhon:01}. They
are characterized by the Arnowitt-Deser-Misner (ADM) mass-energy
$M_\text{ADM}$ and angular momentum $J_\text{ADM}$. The initial
separation is $40$~km ($\sim$$3$ orbits to merger).  The spacetime is
evolved with the Z4c formulation~\cite{Bernuzzi:2009ex}, coupled with
GRHD and a neutrino leakage
scheme~\cite{Galeazzi:2013mia,*Radice:2016dwd}.  We employ the
\texttt{Einstein Toolkit} \cite{Loffler:2011ay} with the
\texttt{CTGamma} spacetime solver and the 
\texttt{WhiskyTHC} GRHD code
\cite{Radice:2012cu,*Radice:2013hxh,*Radice:2013xpa}.  We use the
high-order MP5 reconstruction implemented in \texttt{WhiskyTHC} to
ensure that the effect of numerical dissipation is minimized. The
Courant-Friedrichs-Lewy factor is set to $0.15$ to guarantee the
positivity preserving property of the limiter described in
\cite{Radice:2013xpa}. Dynamical evolutions are carried out with 
linear resolution of $\Delta x=295$~m for a total time of $\sim$$60$~ms
after merger, and with $\Delta x=185$~m for $20$~ms after merger.  Our
grid consists of 6 refinement levels with the coarsest being a cube of
linear extent $1024 M_\odot \simeq 1512\ \mathrm{km}$. To reduce our
computational cost, we impose symmetry across the $xy-$plane and, for
equal mass models, we assume $\pi-$symmetry. Model LS220-135135 is
simulated also without leakage.  The GWs are extracted from the
spin-weighted multipolar decomposition of the Weyl scalar $\Psi_4$
on a sphere placed at $200\ M_\odot \simeq 295\ \mathrm{km}$.

In all simulations but SFHo-144139, we observe the formation of a
HMNS. We define the merger time $t_0$ as the time of waveform peak
amplitude
\cite{Bernuzzi:2015rla}; time periods of $N$~ms after $t_0$
are indicated as $t_{N}$. Figure~\ref{fig:rhomax} shows the evolution
of the maximum rest-mass density $\rho_{\max}(t)$ for all models and
snapshots of the temperature $T$ and rest-mass density $\rho$ in the orbital plane at
representative times for LS220-135135 (with leakage).

During merger, the two NS cores come into contact and merge to a single
core within $\sim$$t_{10}$. $\rho_{\max}$ increases by up to a factor
$1.5$$-$$2$ and oscillates violently. Note that for a given total mass,
stiffer EOS have smaller $\rho_{\max}$. Additionally, the oscillations
in $\rho_{\max}$ have higher amplitude when the configuration is
closer to the prompt collapse threshold and when $\rho_{\max}$ is
larger. 
The evolution from the initial two-core structure 
into a more axisymmetric single-core HMNS is due to hydrodynamic
angular momentum redistribution and dissipation by shock heating and
GW emission~\cite{Shibata:2006nm}. The extreme nonaxisymmetric shape and the
increase in density result in very efficient GW
emission~\cite{Bernuzzi:2015rla}. 

Temperatures as high as $\sim50\, \mathrm{MeV}$ are reached in the
interface between the NSs (Fig.~\ref{fig:rhomax}).  Physically, we
expect these temperatures to be somewhat lower, because at very high
resolutions and when MHD is included, \cite{Kiuchi:2015sga} showed
that a fraction of the shear flow energy created at contact is
converted into magnetic field energy. In our simulations, instead, the
unresolved shear energy is converted into heat by our finite-volume
scheme. This corresponds to a case in which no large-scale dynamo is
activated and the locally generated magnetic field dissipates.

As the merger and the early HMNS evolution proceed, we observe hot
streams of matter being squeezed out of the interface between the two
NSs. Part of this material becomes unbound while the rest forms a
thick torus around the merger remnant. As the two NS cores merge, the core
remains relatively cold, with $T\sim$$10\ \mathrm{MeV}$, while the
temperature peaks at around $\sim$$50\ \mathrm{MeV}$ at densities of
$\sim$$3-5\times 10^{14}\ \mathrm{g}\ \mathrm{cm}^{-3}$. Even at these
lower densities, the EOS is only mildly affected by thermal effects
\cite{kaplan:14}.

The high mass of SFHo-144139, combined with the particularly soft EOS,
results in prompt collapse at merger.  We observe BH
formation within the simulated time also for LS220-1365125,
LS220-135135, LS220-144139, SFHo-135135. It is interesting to note
that LS220 and SFHo have similar cold non-spinning NS maximum masses,
but SFHo HMNSs collapse much more quickly. This is due to their
more compact postmerger configuration, which leads to a more rapid
evolution toward instability \cite{kaplan:14}. We list the remnant BH
masses and spins in Tab.~\ref{tab:bns}. The properties of the
accretion disks will be discussed elsewhere \cite{RichersInPrep}.

\section{GW Energy and Angular Momentum}
The energy radiated in GWs over the entire history of the binary up to
the start of our simulations ($t=0$), is (in $G = c = 1$)
$E_{\mathrm{GW},i} = M - M_\mathrm{ADM}$, where $M = M_A + M_B$ is the
binary gravitational mass at infinite separation.  From the $\Psi_4$
projections we compute the waveform multipoles $h_{\ell m}(t)$ up to
$\ell=\ell_\mathrm{max} = 8$, and, using Eqns.~(15) and (16) of
\cite{Bernuzzi:2012ci}, the energy and angular momentum emitted in GWs
during our simulations, $\Delta E_\text{GW}(t)$ and $\Delta
J_\text{GW}(t)$, respectively.  The total emitted energy over
inspiral, merger, and postmerger evolution to time $t$ is then
$E_\mathrm{GW}(t) = E_{\mathrm{GW},i} + \Delta
E_\text{GW}(t)$. Similarly, the binary angular momentum to time $t$ is
given by $J(t) = J_\text{ADM} - \Delta J_\text{GW}(t)$.  We report
both quantities normalized by $M$ at different times in
Tab.~\ref{tab:bns}.

\begin{figure}[t]
  \includegraphics[width=\columnwidth]{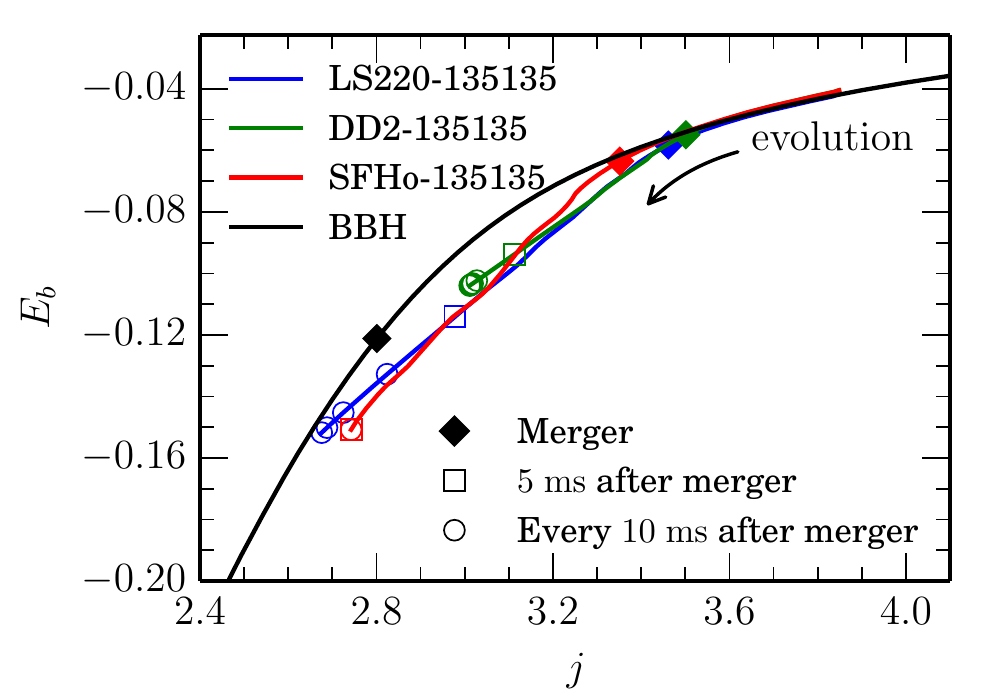}
  \caption{BNS dynamics in terms of gauge-invariant binding energy vs. angular
  momnetum curves. Equal-mass configurations are compared to the
  corresponding nonspinning BH binary. 
  The largest GW luminosity comes from the HMNS, and the overall energy
  emission (relative to the mass) from BNS is in many cases larger than the
  BH inspiral-merger case (excluding ringdown). These features are common to all our simulated BNS.}
  \label{fig:Ej}
\end{figure}

A gauge-invariant way to represent the HMNS GW emission is to consider
binding energy vs. angular momentum curves in analogy to the approach
proposed in~\cite{Damour:2011fu,Bernuzzi:2012ci}.  Working with
quantities per reduced mass, we define $E_b = -E_\text{GW}/(M\nu)$ and
$j=J/(M^2\nu)$ with the symmetric mass ratio $\nu=M_A M_B/M^2 \approx
1/4$.  Representative examples of $E_b(j)$ curves are shown in
Fig.~\ref{fig:Ej}.  The binary evolution starts at large $j$ (large
separations) and at small negative $E_b$, accounting for the energy
radiated over the inspiral until the point our simulations start. GW
emission drives the system to smaller $j$ and lower $E_b$.
Importantly, the largest change of $E_b$ and $j$ (corresponding to the
highest GW luminosity) occurs within $t_{10}$ after
merger. Furthermore, the $E_b(j)$ curves in the HMNS phase are
approximately linear, indicating that the main emission is at an
approximately constant frequency proportional to the derivative
$\partial E_b/\partial j$~\cite{Bernuzzi:2015rla}.

During inspiral and up to merger ($t_0$, diamonds in
Fig.~\ref{fig:Ej}), the BNS typically emits $1.27-1.66\%$ of its
initial mass-energy $M$~\cite{Bernuzzi:2014kca}.  \emph{The energy emission
within $t_{10}$ is up to twice as large as the energy emitted during
the whole inspiral}!  By the end of our simulations ($t_{50}$ or
collapse), the BNS has typically emitted $\sim$$2.18-3.93\,\%$ of $M$
(cf.~Fig.~\ref{fig:Ej}). This fractional energy emission is comparable
to -- or larger than -- that of a nonspinning equal-masses BH binary
inspiral-merger ($\sim$$3$\%), excluding the ringdown ($\sim$$5$\%). However, quasicircular BH binaries
with aligned spins can emit up to $13\%$ of
$M$~\cite{Campanelli:2006fg,Hemberger:2013hsa}; high-energy BH
collisions up to $\sim$$60\%$~\cite{Sperhake:2012me}.  If the HMNS
survives for $t$$>$$t_{20}$, then the GW energy contribution from the
subsequent part of the evolution is negligible.  These
considerations hold also for configurations like LS220-144139, whose
HMNS collapses within $t_{10}$, but obviously not for the prompt
collapse case SFHo-144139 (no HMNS).

Our results show that the details of the above depend 
crucially on EOS and binary mass. In general, for fixed masses, the
stiff DD2 EOS gives the smallest energy emission. For fixed EOS, the
larger the total mass, the larger is the GW energy emission
relative to the total mass. 
However, in the case of a configuration close to the collapse
threshold that collapses soon after merger ($\Delta t\ll t_{10}$),
lower rather than higher masses favor GW energy/angular momentum emission
(cf.\ LS220-144139 vs. LS220-135135 and SFHo-135135
vs. SFHo-136125).

The dimensionless mass-rescaled angular momentum available at merger
is in the range $3.3\lesssim j(t_0)\lesssim 3.6$ ($0.83\lesssim
J(t_0)/M^2\lesssim0.89$); this range is representative of a large
sample of EOS, masses, and mass ratios~\cite{Bernuzzi:2014kca,Bernuzzi:2015rla}.
The GW emission during the early HMNS evolution reduces these values
by $11$$-$$22\,\%$, depending on binary configuration and EOS.  The
late-time value of $J(t)/M^2$ is the largest spin $a_\text{BH}$ that
the remnant BH can have (assuming no disk is produced). For HMNSs that
collapse within $t_{50}$, an upper limit for the 
BH spin parameter is $\max(a_\text{BH})\lesssim 0.7$ ($j
\lesssim 2.8$ for $\nu = 1/4$, cf.~Fig.~\ref{fig:Ej}). The angular
momentum evolution of HMNSs that are stable beyond $t_{50}$ is
expected to be significantly affected by MHD angular
momentum redistribution and breaking and is presently highly
uncertain.

\begin{figure}[t]
  \includegraphics[width=\columnwidth]{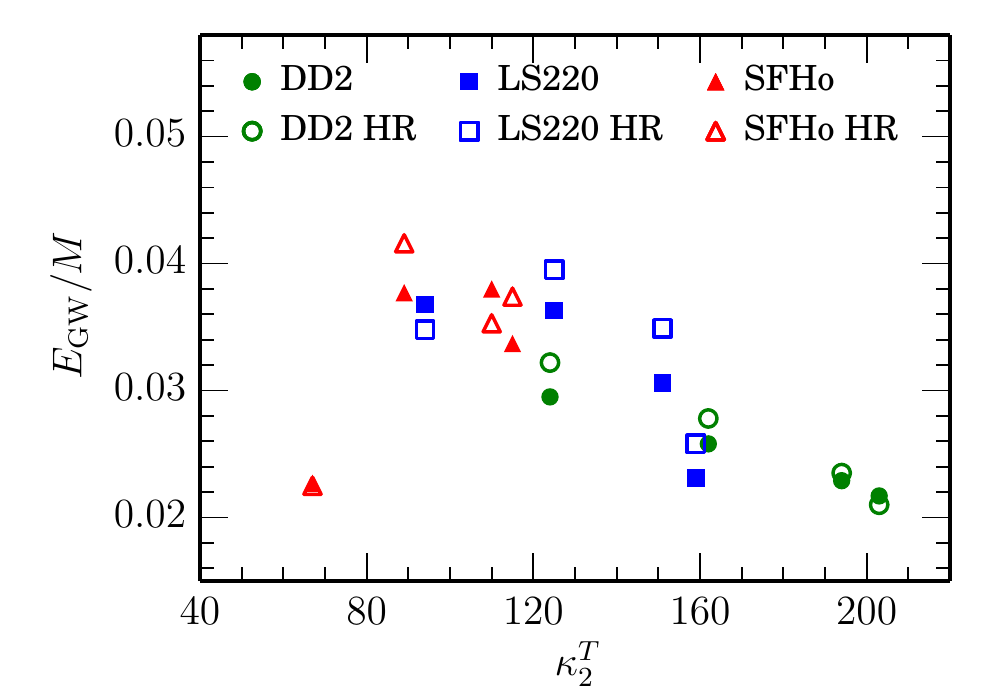}
  \caption{Dimensionless and mass-rescaled GW energy at $t_{20}$ (or
    $t_c$$<$$t_{20}$) as a function of the tidal coupling constant $\kappa^T_2$~\cite{Bernuzzi:2014kca}.}  
  \label{fig:Egw_kappa}
\end{figure}

Runs at higher-resolution (HR) show that our results are robust
and actually conservative: the GW luminosity is typically
underestimated due to numerical dissipation at low resolution.  The
HMNS collapse time $t_c$ can vary by a few milliseconds for
configurations close to the collapse treshold, e.g.\ LS220-144139 has 
$t_c$$\sim$$t_{6}$ for $\Delta x=185$~m runs, while $\sim$$t_{10}$
for $\Delta x=290$~m. The respective $E_\text{GW}(t_{20})$ variation
is, at most, $\lesssim$$10\%$ at HR. However, because a HMNS that
collapses earlier also emits more GWs early on, the timescale of the
main GW emission remains $\sim$$t_{10}$.

\section{Discussion}  
We demonstrate for the first time that,
due to the extreme densities and nonaxisymmetry of the early
postmerger phase, generic BNS 
mergers can reach large GW luminosity corresponding to
$L_\text{GW}\sim6\times10^{55}\,\mathrm{erg\,s}^{-1}$, with typical
emission timescale of $\sim$$t_{10}$ (compare with~\cite{Abbott:2016blz}).
Our results lead us to the conjecture that the maximum postmerger GW
emission efficiency is attained by a configuration in which EOS and
binary mass are such that the HMNS is slighly below the prompt
collapse threshold and supported for $\sim$$t_{10}$. Such
configurations can be identified by investigating the dependence on
the coupling constant for tidal interactions
\cite{Bernuzzi:2015rla}. The latter is defined as
$\kappa^T_2=\kappa^A_2+\kappa^B_2$, with $\kappa^A_2 = 2 k^A_2
\left(X_A/C_A\right)^5 M_B/M_A$, where $C_A$ is the compactness of
star $A$, $X_{A}=M_{A}/M$, and $k_{2}^{A}$ the quadrupolar
dimensionless Love number \cite{Bernuzzi:2012ci}.  Large values of
$\kappa^T_2$ correspond to stiff EOS (large Love numbers) and
individual stars with low compactness, see Tab.~\ref{tab:bns}. The
number $\kappa^T_2$ parametrizes, at leading order, tidal interactions
during the orbital phase and is the key parameter to effectively
characterize merger dynamics and postmerger GW
frequencies~\cite{Bernuzzi:2015rla}. The total GW energy is shown as a
function of $\kappa^T_2$ in Fig.~\ref{fig:Egw_kappa}, which includes
results from high- and low-resolution simulations. These results
suggest that the maximum GW efficiency is obtained for binaries with
$70\lesssim\kappa^T_2\lesssim 150$. This is a narrow range compared with
the $\sim$$10$$-$$500$ range of values that $\kappa^T_2$ may
assume for BNS systems~\cite{Bernuzzi:2014kca}. The efficiency maximum
is caused by the competition between BH formation, occurring earlier
for smaller $\kappa^T_2$, and the GW energy emission decreasing with
increasing $\kappa^T_2$.  A larger $\kappa^T_2$ corresponds to a
larger tidal disruption radius, a less compact postmerger
configuration with a smaller angular frequency, and therefore less
energy loss relative to angular momentum loss.
 
Observational constraints on the EOS could be obtained by
combining a single GW energy measurement 
with the results in Fig.~\ref{fig:Egw_kappa}. 
More simulations and a
more accurate characterization of the relation
$E_\text{GW}(\kappa^T_2)$ are required for this purpose. Most
importantly, observing the large GW luminosities reported here will be
challenging for the Advanced LIGO/Virgo detectors because of the high frequency
($2$$-$$4$~kHz) nature of the emission. The typical horizon distance for a
signal-to-noise ratio $9$ is $\sim$$10$~Mpc for an optimally oriented
source. Unless optimized sensitivity
curves at high-frequencies are developed, the postmerger GW spectrum
will remain a target for third generation
detectors~\cite{Clark:2015zxa}.

Due to the short timescale of the GW emission ($t_{10}$), physical
processes other than hydrodynamics and shock-heating are unlikely to
affect  the emission. For the LS220-135135 case, we have verified that neutrino
cooling does not affect the GW emission in $t_{50}$.
Similarly, MHD effects are expected to influence the GW luminosity
only if they can significantly affect the short-timescale HMNS
dynamics. The magnetorotational instability (MRI) and its ability to
redistribute angular momentum might drive the HMNS to an early
collapse. This can be characterized by an effective viscosity, which
is currently poorly constrained, but simulations of \cite{kiuchi:16priv}
suggest an angular momentum redistribution timescale of
$\mathcal{O}(100)\,\mathrm{ms}$.
Thus, also for the MRI, we expect little influence on the GW
luminosity. Future, very high-resolution MHD simulations are necessary
to further test this
assertion~\cite{Kiuchi:2015sga,Mosta:2015ucs,kiuchi:16priv}.

Finally, our new limit on the spin of the final BH has important
consequences for models of sGRBs relying on the energy deposition by
neutrino pair-annihilation. There, the energy deposition rate depends
strongly on the BH spin \cite{Zalamea:2010ax}. For fixed accretion
rate, the energy deposition by neutrinos from a disk accreting onto a
BH with $a = 0.7$ can be up to a factor $\sim$$100$ times smaller than
for a disk feeding a maximally spinning BH \cite{Zalamea:2010ax}.  Our
limit on $a$ does not significantly constrain sGRB models invoking
magnetic effects, which can easily account for the required energies
even in absence of extremely high BH spin, e.g., \cite{nakar:07}.

Our waveforms are publicly available on {Z}enodo's {NR}-{GW}
{O}pen{D}ata community \cite{bernuzzi_2016_57844}.

\begin{acknowledgements}
We thank B.S.~Sathyaprakash for
triggering this work. This research was partially supported by the
Sherman Fairchild Foundation and by NSF under award Nos.\ CAREER
PHY-1151197, PHY-1404569, and AST-1333520. The simulations were
performed on the Caltech computer Zwicky (NSF PHY-0960291), on NSF
XSEDE (TG-PHY100033), and on NSF/NCSA Blue Waters (NSF PRAC
ACI-1440083).  LR and PM were supported by NASA Einstein Postdoctoral
Fellowships under grant numbers PF3-140114 and PF5-160140,
respectively.
\end{acknowledgements}

\bibliographystyle{apsrev} 
\bibliography{refs20160713}

\end{document}